\documentclass[final,5p,times,twocolumn,sort&compress]{elsarticle}

\usepackage{amssymb}
\usepackage{xcolor}
\usepackage{siunitx}
\DeclareSIUnit{\molar}{M}

\begin{document}

\begin{frontmatter}



\title{Self-Lubricating Drops}
\date{}

\author[label1]{Huanshu Tan}
 \ead{tanhs@sustech.edu.cn}
\affiliation[label1]{organization={Center for Complex Flows and Soft Matter Research \&
Department of Mechanics and Aerospace Engineering, Southern University of Science and
Technology},
            city={Shenzhen},
            postcode={518055}, 
            state={Guangdong},
            country={P.R. China}}

           \author[label2]{Detlef Lohse}
 \ead{d.lohse@utwente.nl}
\affiliation[label2]{organization={Physics of Fluids Group, Max Planck Center Twente for Complex Fluid Dynamics, Department of Science and Technology, and J. M. Burgers Centre for Fluid Dynamics, University of Twente},
city={Enschede},
            postcode={7500 AE },
            country={The Netherlands}}  
            
 \author[label3]{Xuehua Zhang}
 \ead{xuehua.zhang@ualberta.ca}
\affiliation[label3]{organization={Department of Chemical \& Materials Engineering, University of Alberta},
            city={Edmonton},
            postcode={T6G1H9}, 
            state={Alberta},
            country={Canada}}
            
\begin{abstract}

Over the past decade, there has been a growing interest in the study of multicomponent drops.
These drops exhibit unique phenomena, as the interplay between hydrodynamics and the evolving physicochemical properties of the mixture gives rise to distinct and often unregulated behaviors.
Of particular interest is the complex dynamic behavior of the drop contact line, which can display self-lubrication effect.
The presence of a slipping contact line in self-lubricating multicomponent drops can suppress the coffee-stain effect, conferring valuable technological applications.
This review will explain the current understanding of the self-lubrication effect of drops, and cover an analysis of fundamental concepts and recent advances in colloidal assembly.  The potential applications of self-lubricating drops across different fields will also be highlighted. 

\end{abstract}




\end{frontmatter}


\section{Introduction}
\label{}


The contact line behavior of the drying drops on surfaces has garnered substantial interest due to its significant relevance to various practical applications, including surface patterning~\cite{deegan1997capillary, yunker2011suppression}, inkjet printing~\cite{de2004inkjet, lohse2022fundamental}, healthcare diagnostics~\cite{sobac2011structural}, and advanced material fabrication~\cite{dabodiya2022ultrasensitive}.
In a groundbreaking study, Deegan~\textit{et al.} demonstrate the principle for the ``coffee-ring" or ``coffee-stain" effect~\cite{deegan1997capillary,deegan2000contact}, 
as illustrated in Figure~\ref{fig:fig1}a.
The contact line dynamics of multicomponent sessile drops are characterized by complex behavior resulting from the interplay between physiochemical properties and interaction among the different components~{\cite{marin2011order,lohse2020physicochemical,wang2022wetting,tan2022evaporation,al2023binary}. 
These characteristics include changes with drop composition in surface tension~\cite{sefiane2003experimental,karpitschka2017marangoni}, in gravitational effect~\cite{li2019gravitational}, in chemical reactivity~\cite{kumar2007review,eustathopoulos2016role},  and in freezing and boiling behaviors~\cite{marin2014universality,lyu2021explosive}.
Understanding the dynamics of the contact line is crucial for a variety of applications, and comprehensive reviews of self-cleaning surface, coating technology, and related topics can be found in the literature~\cite{lohse2022fundamental,wang2022wetting,blossey2003self,ganesh2011review,giorgiutti2018drying,dalawai2020recent,sarkin2020review,gelderblom2022evaporation}.

In recent years, a novel concept known as ``self-lubrication" of multi-component sessile drops~\cite{tan2016evaporation,tan2019porous} has gained increasing attention.  
Referred to as self-lubricating drops, this emerging drops exhibit a distinct behavior in contact line dynamics, allowing the suppression of the coffee-stain effect.
Studies have demonstrated the potential of self-lubricating drops in various state-of-art applications for fabrication of nanomaterials, especially via colloidal assembly~\cite{tan2019porous,thayyil2021particle,koshkina2021surface,li2022self}.

In this article, we begin with an introduction to the self-lubrication of multicomponent sessile drops and its underlying mechanism. 
We then delve into the discussion of the dynamic processes involved, including liquid-liquid phase separation, compositional evolution, Marangoni interfacial flow, and surface property influence.  
Expanding on these fundamental concepts, we present a concise overview of recent advancements in self-lubricating drops,
including supraparticle fabrication, the effects of particle properties on the fabrication process, the application in sensing and detection, and exploration of different liquid drop systems exhibiting self-lubricating behavior.
Finally, we conclude our review by posing several open questions and offering our current opinions on relevant topics.

\section{Self-lubrication of multicomponent sessile drops}
\label{}

\begin{figure*}[h]
\centering
\includegraphics[width=0.9\linewidth]{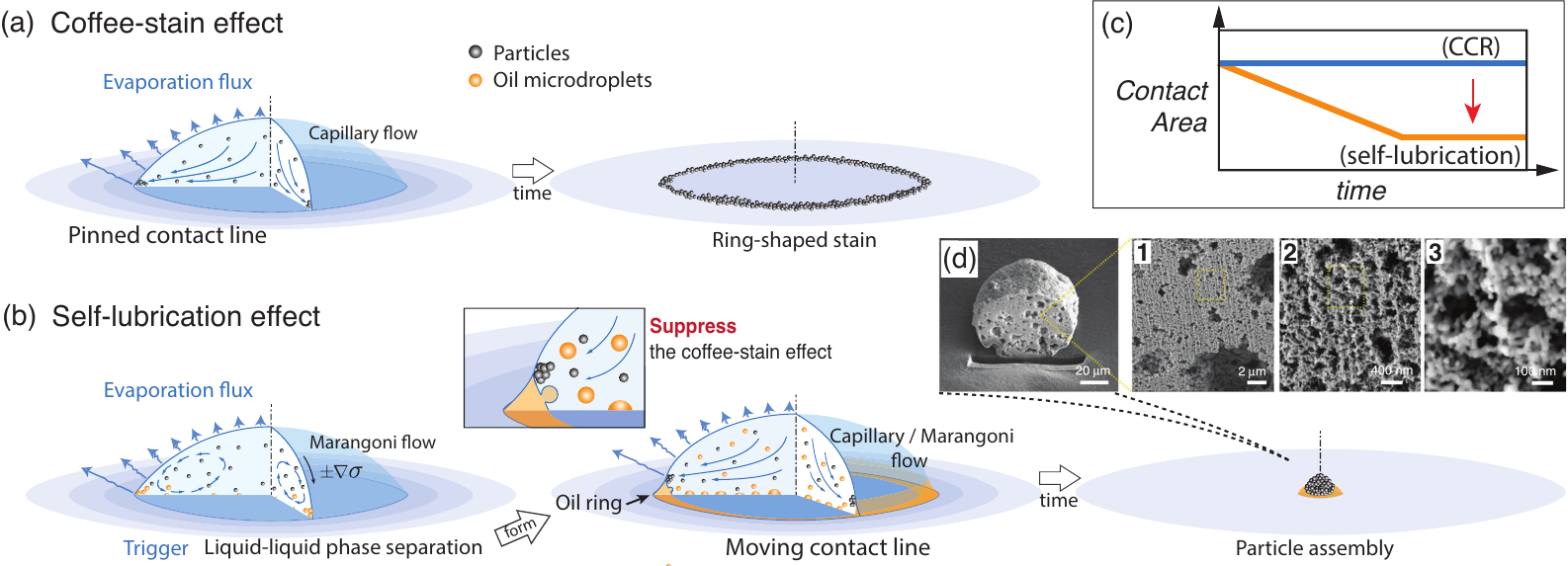}
\caption{Suppression of the ``coffee-stain'' effect by the ``self-lubrication'' effect. 
Illustrations of (a) the ``coffee-stain" effect, where the capillary flow and pinned contact jointly lead to a ring shaped stain after the drop evaporation, and 
(b) the ``self-lubrication'' effect, where an oil ring spontaneously forms at the drop contact line and suppresses the ``coffee-stain'' effect, thereby resulting in the assembling of particles, i.e. forming a supraparticle.
(c) Distinguished evolutions of the drop contact area between the ``coffee-stain" effect (CCR mode) and the ``self-lubrication'' effect.
(d) SEM photographs of the generated supraparticle by drying a self-lubricating sessile drop ladened with TiO$_2$ nanoparticles, showing the fractal-like interior structure. 
Panel (d) is adapted with permission from ref~\cite{tan2019porous}.
}
\label{fig:fig1}
\end{figure*}

\subsection{Dynamic contact line modes}

In this subsection, we will provide a brief review of the typical dynamic modes exhibited by evaporating sessile drops, which will serve as a foundation for introducing the concept of self-lubrication.

The evaporation of a liquid drop on a solid surface can occur within different dynamics contact line modes~\cite{gelderblom2022evaporation,picknett1977evaporation,bonn2009wetting,dietrich2015stick}.
Specifically, the drop contact line can slide freely with a constant contact angle, in the constant contact angle (CCA) mode. 
Whereas, in the constant contact radius (CCR) mode, the drop maintains a constant contact radius during the evaporation process. 
As illustrated in Figure~\ref{fig:fig1}a, CCR mode is a characteristic of the coffee-stain effect, essential for the development of the capillary flow that brings ladened particles (or solid contaminants) towards the pinned contact line. 
The particle accumulation consolidates the sticky contact line and CCR mode, leaving a ring-shaped stain on the surface after the drop evaporation~\cite{deegan1997capillary,deegan2000contact}.
Thus, the behavior of the contact line of the drop can significantly influence the dynamics of the evaporation process. 
 
However, evaporation of a multicomponent drop can exhibit enhanced mobility of the contact line through the so-called ``self-lubrication effect"~\cite{tan2016evaporation, tan2019porous}.

\subsection{What is the ``self-lubrication" effect?} 
The term ``self-lubrication" refers to the phenomenon that occurs in evaporating multicomponent sessile drops~\cite{tan2019porous}, as illustrated in Figure~\ref{fig:fig1}b.
The drop evaporation triggers the \textit{spontaneous} formation of an immiscible liquid ring at the drop contact line.
The emerging oil ring promotes the sliding contact line of the evaporating drop.
For example, an oil ring around a water-rich drop serves to prevent the accumulation of suspended particles near the gas-liquid-solid three-phase contact line (surface property effects refer to Sec.~\ref{sec:surfaceproperty}).
There is no landing particles at the contact line to consolidate CCR mode.
Instead there is an effective suppression of the ``coffee-stain" effect. 
As illustrated in Figure~\ref{fig:fig1}c, the contact area size of the residue from the evaporated drop is consequently reduced.
It is remarkable that, without presetting precursor films or utilizing a superhydrophobic substrates, the self-lubrication effect  improves the mobility of the sessile drop, leading to a reduction in the contact area.

In presence of nanoparticles in the self-lubricating drop, the nanoparticles self-assemble into a three dimensional macroscopic structure, known as a supraparticle.
Because of their unique structural features and high porosity~\cite{tan2019porous} (Fig.~\ref{fig:fig1}d), supraparticles are attractive and can facilitate a range of applications~\cite{wintzheimer2021supraparticles, guo2022supraparticle, hu2022multi, li2022conformally, jo2022multimodal}, including catalysis, purification, drug delivery, and energy storage. A brief discussion of the recent advances is given in sections~\ref{sec:s1-1} and ~\ref{sec:s1-2}.

\subsection{Spontaneous formation of the lubricating oil ring}
The lubricating oil ring forms spontaneously from the liquid-liquid phase separation, triggered by the preferential evaporation across the multicomponent drop surface~\cite{tan2016evaporation}.
The subsequent merging of the separated oil component at the drop contact line creates the continuous oil ring.

Typically, the self-lubricating multicomponent drops comprise \textit{a minimum of three constituents}, one cosolvent and two immiscible fluids that may undergo phase separation from reducing the cosolvent function~\cite{lu2017dissolution,moerman2018emulsion,tan2019microdroplet}. 
The cosolvent augments the degree of miscibility between the two immiscible fluids.
For example, water and oil are soluble in the mixture, by introducing ethanol in certain quantities~\cite{tan2016evaporation,tan2017self}.
However, in situations where the concentration of the cosolvent molecules fails to achieve thermodynamic mediation between the two immiscible components, new liquid-liquid interfaces starts to form, leading to liquid-liquid phase separation.
This process of phase separation can be triggered spontaneously by the evaporation (dissolution) of the self-lubricating multicomponent drops, owing to the high volatilities (solubility) of the cosolvent~\cite{guo2021non,chao2022liquid,may2022phase}.
Furthermore, it is important to note that the composition of the drop plays a significant role in phase separation, which will be discussed in detail in section~\ref{sec:sec-composition}.

\begin{figure*}[t]
\centering
\includegraphics[width=0.9\linewidth]{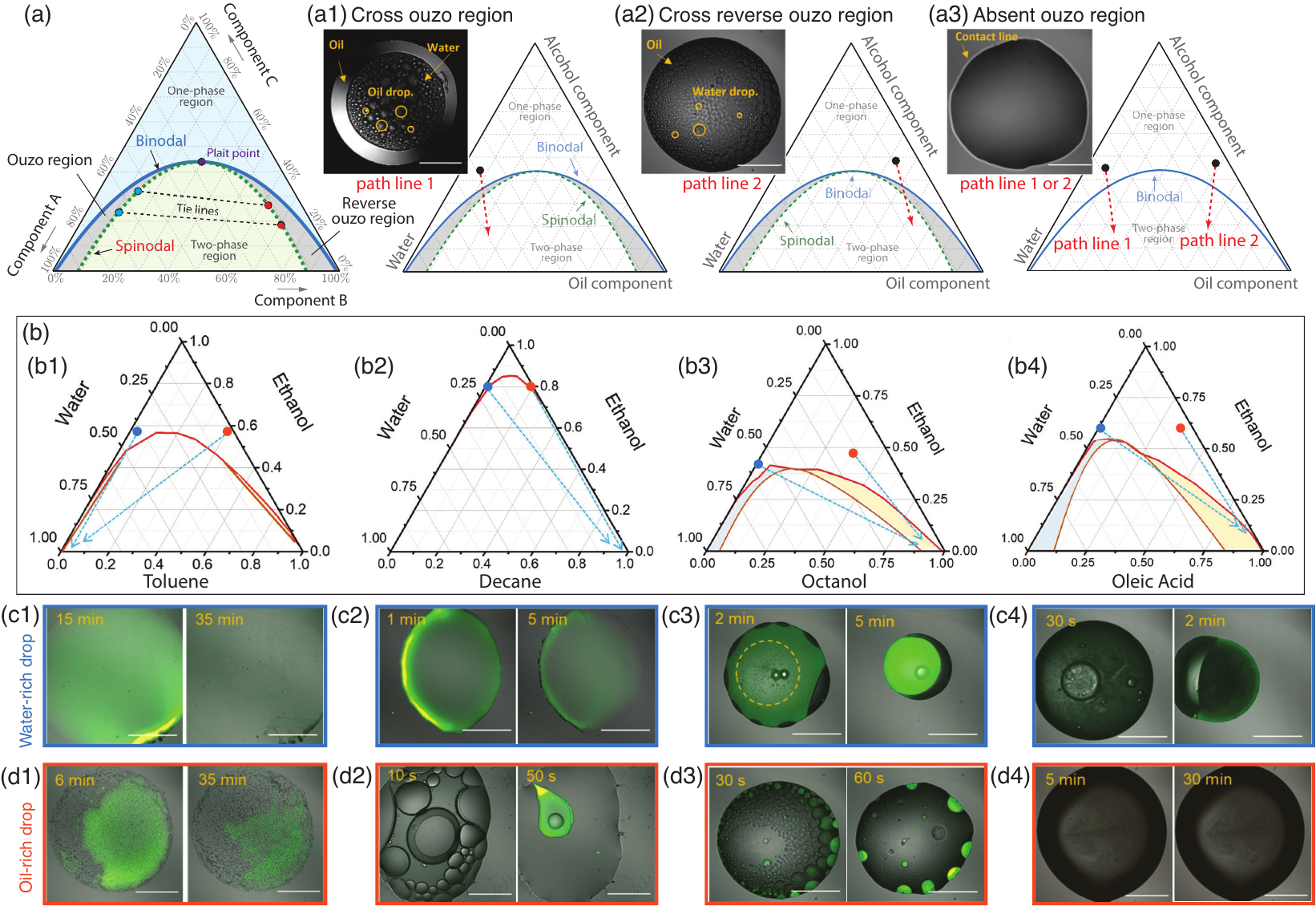}
\caption{Phase diagram and the role of composition path line. 
(a) Ternary phase diagram with ouzo and reverse-ouzo regions. (a1-a3) For the same ternary liquid system, different path lines (dashed red lines) of drop's compositional evolution result in three different configurations of the separated liquid phases. 
(b) Ternary phase diagrams of water, ethanol and one of the four different oils, (b1) toluene, (b2) decade, (b3) octanol, and (b4) oleic acid. 
Distinguished confocal images of the evaporation process of a C-dot ternary drop in water-rich compositions (c1-c4) and oil-rich compositions (d1-4), respectively corresponding to the blue dots and the orange dots in the four different ternary liquid systems (b1-b4). 
The scale bars in (a1-a3), (c1-c4) and (d1-d4) are 100 \SI{}{\mu m}.
Panels (a1-a3), (b1-b4), (c1-c4), and (d1-d4) are adapted with permission from ref~\cite{li2022self}.
}
\label{fig:fig2}
\end{figure*}

The evaporation-triggered phase separation includes preferred heterogeneous nucleation at the substrate~\cite{zhang2015formation} or in the bulk.
The former directly leads the formation of surface microdroplets of the separated component fluid, including these oil microdroplets at the drop contact line, while numerous oil microdroplets nucleate in the bulk.
As illustrated in Figure~\ref{fig:fig1}b, the growing and merging-up of oil microdroplets  (orange dots) along the contact line form the self-lubricating oil ring.
The convective flow in the drop drives the suspended oil microdroplets to move to the contact line contributing to the oil ring formation.
\\

Having introduced the concept of self-lubrication, it becomes evident that the complex behavior of the multicomponent drop is a result of the interplay between the physicochemical properties of the components and the micro-hydrodynamics induced by evaporation~\cite{lohse2020physicochemical,moerman2018emulsion,chao2022liquid,tan2018evaporation,baumgartner2022marangoni}. 
It is the coupling that affects the liquid-vapor equilibrium at the drop surface and the liquid-liquid equilibrium within the solution~\cite{tan2022evaporation}.

\begin{figure*}[h]
\centering
\includegraphics[width=0.9\linewidth]{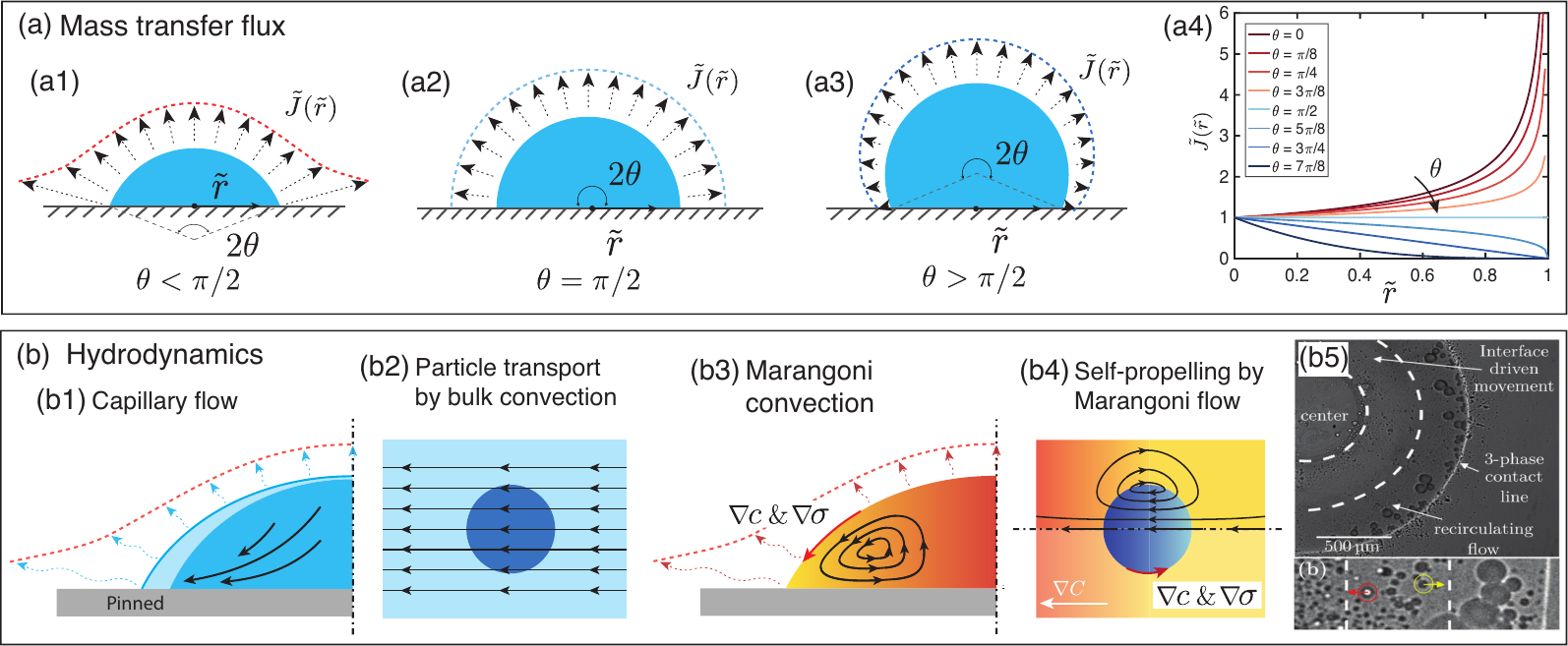}
\caption{Underlying fluid dynamics~\cite{deegan1997capillary,deegan2000contact,leal2007advanced,cazabat2010,tan2016evaporation,maass2016swimming,tan2018evaporation,tan2021two,may2022phase}.
(a) Three typical distributions of the diffusive (evaporation/dissolution) flux at the surface of sessile drop (a1-a3), and the dimensional profiles of the diffusive flux for three different contact angles.
(b) Four typical flow motions that affect the dynamic behaviors of the drop, including (b1) capillary flow, (b2) particle (microdroplet) transport by bulk flow within the sessile drop, (b3) Marangoni surface flow due to the compositional gradient along the sessile drop surface, and (b4) self-migration of particle (microdroplet) driven by the  gradient inside the sessile drop.
(b5) Experimental images that present the simultaneous occurrence of the bulk-transport and self-propulsion of microdroplets in a self-lubricating drop. 
Panel (b5) is adapted with permission from ref~\cite{may2022phase}.
}
\label{fig:fig3}
\end{figure*}

\section{Underlying dynamics and the impacts on self-lubricating drops}
\label{}

To gain insight into self-lubricating drops, the ternary phase diagram can be utilized to quantify the liquid-liquid phase separation. This section discusses solution disequilibrium, interfacial flows, particle interactions, and these dynamics are crucial for understanding the behavior of self-lubricating drops and their advancements in diverse applications.

\subsection{Phase-diagram-quantified liquid-liquid phase separation}
\label{sec:sec-composition}

The ternary phase diagram, as shown in Figure~\ref{fig:fig2}a, is a graphical representation of a three-component system, where each component is represented by a corner of an equilateral triangle. The positions within the triangle correspond to all possible mixture compositions of the three components. The maximal fraction of a component is at its corresponding corner, representing 100\%, while the minimal fraction is at the opposite edge, representing 0\%. 
Using a ternary phase diagram is convenient to characterize the compositional evolution, since a minimum of three components is necessary for the formation of lubricating oil ring~\cite{lohse2020physicochemical,li2022self,guo2021non,tan2018evaporation,lu2015solvent}.

In the ternary phase diagram, where the immiscible components A and B represent water and oil respectively, and the component C is a cosolvent such as an alcohol.
The diagram is divided into two distinct regions by the so-called \textit{binodal curve}.
The region outside the binodal curve, the blue region in Figure~\ref{fig:fig2}a, is the one-phase region. In this region, the concentration of the cosolvent, i.e. the alcohol component, is sufficiently high to render a homogenous solution of water, oil, and alcohol.
However, within the binodal curve (green region), the concentration of alcohol is insufficient to facilitate miscibility between the water and oil components, leading to a ternary isothermal liquid-liquid phase separation, i.e. a two-phase region.
Upon phase separation, the mixture separates into two phases, a water-rich liquid phase and an oil-rich liquid phase.
Each coexisting composition pair of the two phases are connected by a tie line, which is governed by a thermodynamic condition of the equality of the chemical potential~\cite{tan2019microdroplet,cussler2009diffusion,chu2016dissolution}.
This liquid-liquid equilibrium condition ensures that there is no net transfer of matter or energy between two coexisting liquid phases.

The phenomenon of spontaneous emulsification can manifest when the mixture composition lies within the region demarcated by the binodal and spinodal curves.
This zone as depicted by the grey regions in Figure~\ref{fig:fig2}a, is referred to as the ouzo region~\cite{vitale2003,ganachaud2005nanoparticles}. 
The spinodal curve is defined by the condition whereby the second derivative of Gibbs free energy equals zero~\cite{solans2016spontaneous}.
Within the ouzo region, the mixture liquid state is under non-equilibrium conditions and has a positive derivative of Gibbs free energy.
Therefore, minor concentration fluctuations can induce component nucleation, resulting in the formation of liquid phase at a microscopic length scale.
This thermodynamic process leads to the creation of numerous sub-micro or microdroplets that are homogeneously sized and stable, with minimal energy input required~\cite{vratsanos2023ouzo}.

\subsection{Compositional-evolution-determined disequilibrium}

A noteworthy characteristic of self-lubricating drops is its transformation from a miscible drop to a two-phase or even a multi-phase drop system.
The variation of the drop composition leads to solution disequilibrium and the self-formation of an oil ring at the drop's contact line that acts as a lubricant~\cite{tan2016evaporation,tan2019porous}. 
The \textit{path line} is defined as a curve that connects a series of composition points in the ternary diagram, each point representing the composition of the mixture at a specific stage of the process.
Thus, path lines can visualize how the composition of the multicomponent drop changes over time as it undergoes a specific set of conditions~\cite{kenneth1972,tan2019microdroplet}.

During the process of evaporation (dissolution, or reaction), the composition of the self-lubricating drop undergoes changes,  corresponding to an evaporative path line in the ternary phase diagram, as exemplified by the red dashed arrows in Figures~\ref{fig:fig2}-(a1-a3).
Li~\textit{et al.}  reported the distinctive liquid-liquid phase separation phenomena exhibited by self-lubricating drops through different evaporative path lines~\cite{li2022self}. 
When the path line crosses the ouzo region, oil microdroplets form in the water, as shown in the insert of Figure~\ref{fig:fig2}-a1. 
Conversely, when the path line traverses the reverse ouzo region, water microdroplets appears in the oil, as indicated with the orange circles in the insert of Figure~\ref{fig:fig2}-a2.
The control experiment confirmed that no microdroplets form when the ternary liquid system lacks ouzo and reverse ouzo regions, as indicated with the orange circles in the insert of Figure~\ref{fig:fig2}-a3.
Li~\textit{et al.} performed a comprehensive investigation into the self-lubricating drops those containing carbon-nanodots in four distinct oil-water-ethanol ternary systems~\cite{li2022self}.
The corresponding ternary phase diagrams are presented in Figures~\ref{fig:fig2}(b1-b4).
The differing vapor pressure of toluene (3.8 kPa), decane (0.17 kPa), octane (0.0087 kPa), and oleic acid (7.2$\times10^{-8}$ kPa) resulted in a shift in the preference order among the oil, water (3.2 kPa), and ethanol (7.9 kPa) components.
The application of confocal imaging techniques elucidated the influence of the physical properties of the oil component on the self-lubrication process, as presented in Figures~\ref{fig:fig2}c (water-rich drops) and \ref{fig:fig2}d (oil-rich drops).

\subsection{Diffusion-induced hydrodynamics}

In a stationary and diffusion controlled evaporation, free spherical drops in a quiescent open air have uniform diffusive fluxes at their surface,  
however, for sessile drops on a substrate, the situation is different~\cite{cazabat2010}, as illustrated in Figures~\ref{fig:fig3}a1-\ref{fig:fig3}a3. 
The diffusive flux at the drop surface is not uniform any more, except the case that the drop contact angle is 90°.
For the sessile drops with sizes smaller than the capillary length $\ell_c = \sqrt{\sigma/(\rho g)}$ (surface tension $\sigma$, density $\rho$, and the gravitational acceleration $g$), the drops hold spherical-cap shapes, as the surface tension dominates over the gravity.
Given the additional assumption of quasi-steady, the diffusive flux for various contact angle $\theta$ expresses as $J(r) \sim J_0 (1-r/R)^{-\frac{\pi-2\theta}{2(\pi-\theta)}}$~\cite{deegan2000contact,cazabat2010}.
Figure~\ref{fig:fig3}a4 displays the profile of the dimensional diffusive flux for various contact angles, where $\tilde{J}$ and $r$ are normalized by $J_0$ and $R$ respectively.
$J_0$ indicates the diffusive flux at the apex, where the substrate influence is negligible.
For a flat sessile drop ($\theta<$90°), the mass loss near the contact line is faster than other places, as sketched in Figure~\ref{fig:fig3}b1.
Thus, the pinned contact line consequently induces liquid replenishment towards the contact line, i.e. the so-called capillary flow~\cite{deegan1997capillary}.
This rising bulk flow transports the laden particle in the drop (Fig.~\ref{fig:fig3}b2), changing the particle distribution, such as the coffee-stain effect.

In the case of self-lubricating drops of multicomponent liquid, as illustrated in Figure~\ref{fig:fig3}b3, the non-uniform diffusive flux of the volatile component induces the compositional gradients $\nabla c$ and the consequent surface tension gradients $\nabla \sigma$ at the drop surface.
Thus, the interfacial flows arise, known as the solutal Marangoni effect.
The Marangoni convection drives the laden particles and the nucleated microdroplets intensively within the drop.
To gain insight into the hydrodynamics, our theoretical model is based on simplified Navier-Stokes equations for flow, convection-diffusion equations for composition and temperature, Fick's law for diffusion, and multicomponent evaporation model for the mass transfer at the drop surface~\cite{tan2016evaporation,Diddens2017c,Diddens2017b}. These models account for the composition-dependent fluid properties and the Marangoni stress that arises from interfacial tension gradients, and a deeper understanding of the hydrodynamics and transport phenomena within self-lubricating drops can be achieved. It is important to note that while this modeling approach is valuable, it does not account for the evolving drop configuration or liquid-liquid phase separation.

Beside the convective transport, recent researches also reported the self-propulsion of microdroplets inside the self-lubricating drop~\cite{guo2021non,may2022phase}.
The suspended microdroplets in the compositional gradient field also experience the solutal Marangoni effect at the interface~\cite{leal2007advanced,maass2016swimming,tan2021two,zeng2021periodic}, as sketched in  Figure~\ref{fig:fig3}b4.
Hence, the interfacial flows propel the suspended microdroplets in the opposite direction,  similar to that a swimmer pushes the surrounding water backward to move ahead.
May~\textit{et al.} investigated this self-propelling phenomenon in a self-lubricating drop (Fig.~\ref{fig:fig3}b5), and provided the explanation of the mechanism~\cite{may2022phase}.

Based on the discussion of this subsection, it is evident that the above four hydrodynamics -- capillary flow, convective transport of particles, Marangoni flow, and self-propelling of particles -- coexist and interplay, affecting the dynamic behaviors of the self-lubricating drops.

\subsection{Surface property influencing particle interactions}
\label{sec:surfaceproperty}

In a particle-laden self-lubricating drop, the surface properties of colloidal particle play a role in influencing inter-particle interactions and the interactions with various interfaces, including substrates, the drop liquid-gas surface, and the aqueous-oil interfaces. 
In the self-lubricating drops with the presence of nanoparticles, pickering microdroplets (picking emulsions) are formed resulting from the covering of the nanoparticles over the nucleated microdroplets, and their stability and structure depend on the particle-microdroplet interaction~\cite{binks2000influence,goubault2021effect}. 
Manipulating these interactions is essential in governing the self-lubricating process and supraparticle fabrication~\cite{ thayyil2021particle, koshkina2021surface}.

Nanoparticle assembly in the self-lubricating drop results from the competition between a variety of intermolecular and interparticle forces, including Van der Waals forces, magnetic and electrostatic forces, steric forces, depletion forces, hydrodynamic forces~\cite{thayyil2021particle,koshkina2021surface,min2008role}.
Van der Waals forces contribute to the overall interaction energy, while external forces, whether repulsive or attractive forces, allows for delicate manipulation of the assembly and stability of laden particles~\cite{min2008role}. 
Additionally, capillary forces, convective forces, friction and lubrication forces arising from fluid flow and shear have a notable impact on the dynamics of the laden particles~\cite{marin2011order,deegan1997capillary, yunker2011suppression}.

Understanding and manipulating these interactions enable control over the inner structure of particle assemblies, stabilization of pickering microdroplets, and optimize the properties of the generated supraparticles. This opens opportunities,  by applying self-lubricating drops,  for tailoring and engineering systems with desired properties and functionalities.

\begin{figure*}[h]
\centering
\includegraphics[width=0.9\linewidth]{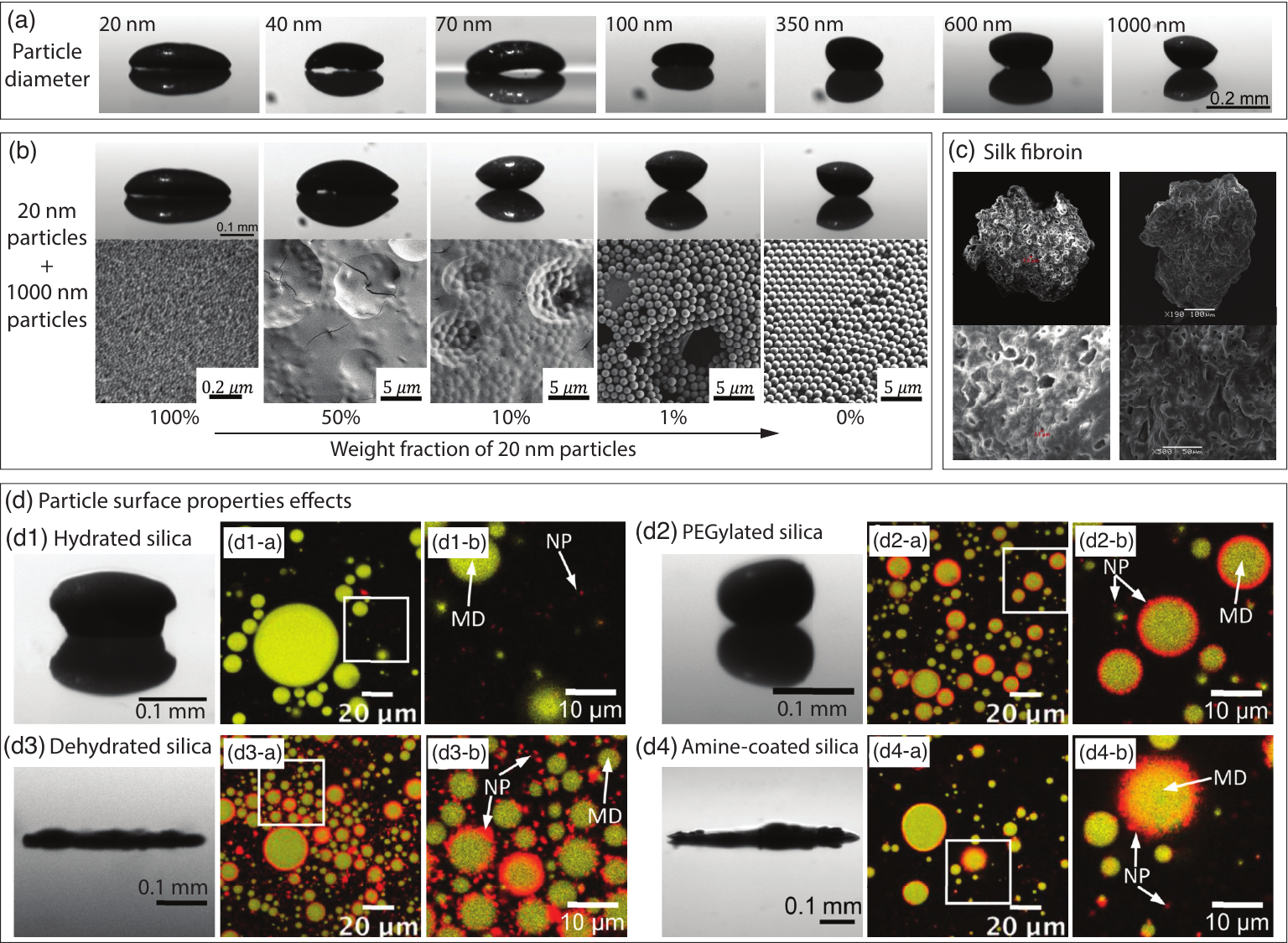}
\caption{Effects of particle surface property on the  assembly process.
(a) Side view of submillimeter-sized supraparticles with different shapes obtained by drying self-lubricating colloidal drops containing silica particles with different sizes.
(b) Varying the fraction of two types of particles with different sizes changes the final shape of the fabricated supraparticle.
Panel (a) and (b) are adapted with permission from ref~\cite{thayyil2021particle}.
(c) Silk fibroin supraparticle fabricated by self-lubricating drops.
Panel (c) is adapted with permission from ref~\cite{lamb2021silk}.
(d) Four different supraparticle shapes (d1-d4) obtained by utilizing four silica particles with different surface properties, as the interplay between the particles (red) and the separated oil component (yellow) affects the particle assembly process (d1-d4).
Panel (d) is adapted with permission from ref~\cite{koshkina2021surface}.
}
\label{fig:fig4}
\end{figure*}

\begin{figure*}[h]
\centering
\includegraphics[width=0.9\linewidth]{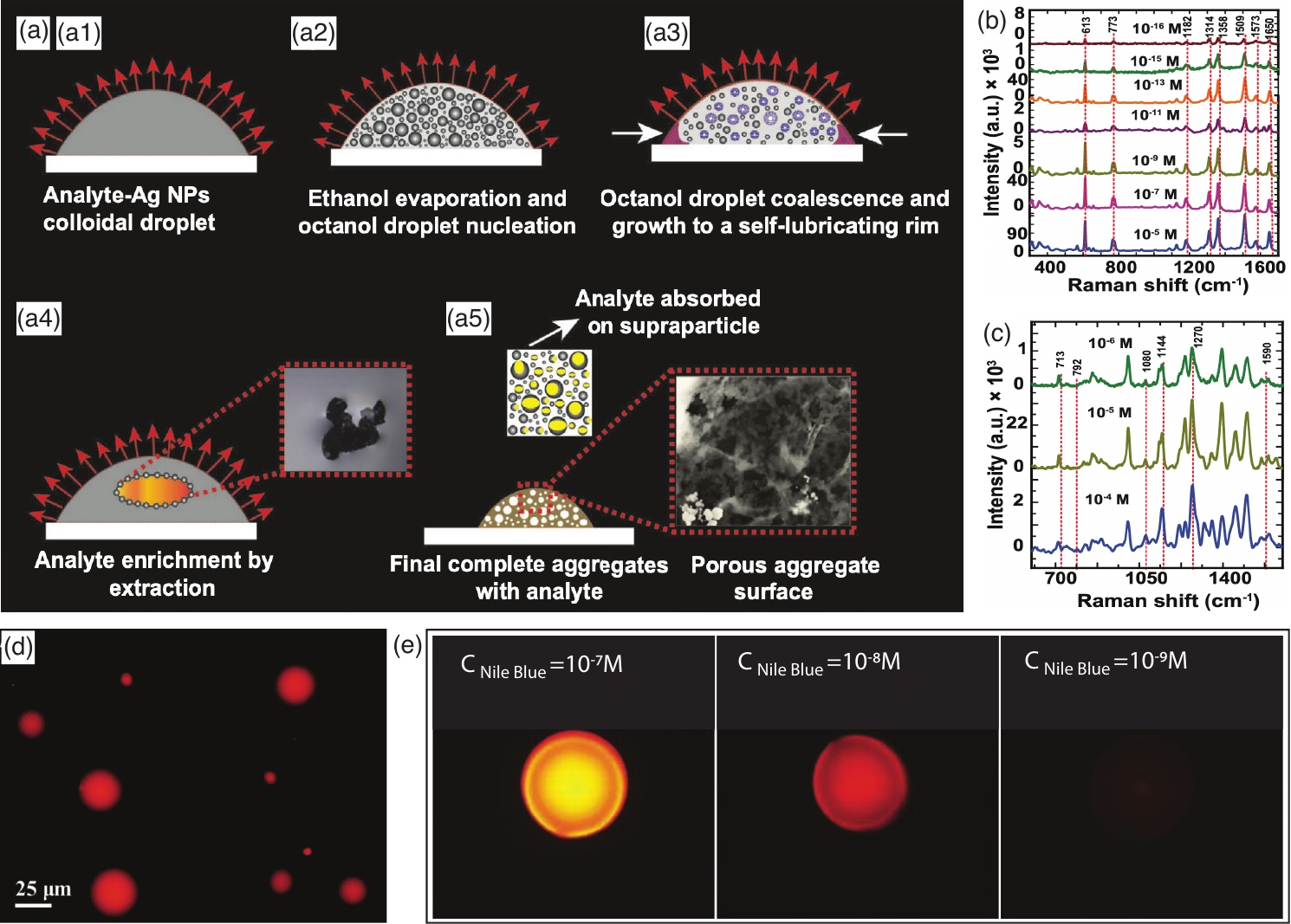}
\caption{Application of self-lubricating drops for sensing and detection.
(a) A schematic representation of applying self-lubricating drops to generate porous Ag-analyte supraparticles, which are used for surface-enhanced Raman spectroscopy detection.
The result shows that the surface-enhanced Raman spectroscopy spectra  can achieve a low detection limit of (b) $10^{-16}$\SI{}{\molar} for R6G detection, and (c) $10^{-6}$\SI{}{\molar} for triclosan detection.
Pannels (a)-(c) are adapted with permission from ref~\cite{dabodiya2022ultrasensitive}.
(d) Nano-extraction enables fluorescence detection of Nile Blue at a concentration that is below the fluorescence detection limit.
(e) Fluorescence images of octanol nanodroplets labeled with different initial concentrations of Nile Blue from $10^{-9}$\SI{}{\molar} to $10^{-7}$\SI{}{\molar} in water.
Pannels (d) and (e) are adapted with permission from ref~\cite{qian2020one}.
}
\label{fig:fig5}
\end{figure*}

\section{Potential applications of the self-lubricating drops}
\label{sec:s1}

The self-lubricating drops have drawn people's attention and being applied in different contexts, particularly in particle assembly. This section presents a review of the recent advances of these approaches, through which we intend to highlight the diversity of the self-lubricating drops and point out open questions in the following section.

\subsection{Particle assembly by self-lubricating drops}
\label{sec:s1-1}
 
The assembly of colloidal particles by drying sessile drops of colloidal suspension is a flexible approach for the fabrication of supraparticles.
However, the coffee-stain effect impedes the assembly of the particles.
Super liquid-repellant substrates have been used to reduce the initial contact area~\cite{rastogi2008,marin2012building,wooh2015synthesis}, thereby weakening the coffee-stain effect. 
But these substrates are fragile, especially in the procedure of supraparticle detachment.
Drying colloidal drop above immiscible liquids can avoid these problems to certain extent, but either the cheerios effect~\cite{vella2005cheerios} or the liquid film replacement cause other issues~\cite{millman2005anisotropic,jiao2022self}.
Therefore, attention is paid to the self-lubricating drops for supraparticle fabrication~\cite{tan2019porous}. 

The supraparticles fabricated by self-lubricating drops have high porosities and tunable shapes.
The liquid-liquid phase separation from numerous oil microdroplets suspending within the colloidal sessile drop.
During the particle assembly, the nucleated oil microdroplets act as templates, resulting in multi-scale, fractal-like structures inside the supraparticle (Fig.~\ref{fig:fig1}d), and the porosity up to 90\%~\cite{tan2019porous}. 
Either spherical or non-spherical shaped supraparticles can be fabricated by adjusting the initial fraction of oil component and the particle of the drop solution.
In the case of the self-lubricating drop having a relatively high oil-to-particle ratio, the particle assembly process in the later stage involves the suspension water drop completely floating on the oil phase. 
This leads to the formation of a spherical shaped supraparticle. 
However, if the oil content is not high enough to completely levitate the aqueous phase, the developing suspension drop undergoes deformation and takes a non-spherical shape~\cite{tan2019porous}.
These unique features of the supraparticles make the particle assembly by self-lubricating drop as an attractive and potential technique.

Raju~\textit{et al.} investigated the particle size effects on the non-spherical shape of the fabricated supraparticles in self-lubricating ternary drops~\cite{thayyil2021particle}.
By using spherical silica particles from 20 to 1000 nm, the authors fabricated pancake-like and American-football-like supraparticles, as displayed in Figure~\ref{fig:fig4}a.
The \textit{in situ} measurements reveal that the supraparticle formation proceeds via the formation of a flexible silica-shell at the contracting oil ring.
The particle size affects the time when the shell-like structure ceases to shrink and loses its flexibility.
With the big particles, the shell ceases deforming by the oil ring at a later ceasing time than the small particles, resulting the American-football-like supraparticles.
The same tendency maintains by increasing the large particle fraction in the mixtures of small and large particles, as presented in Figure~\ref{fig:fig4}b.

Since the hydrophobicity of particles and electrostatic effects can affect the interaction between the oil microdroplets and the colloidal particles, Koshkina~\textit{et al.} used silica particles as a model system to study the particle surface property effect~\cite{koshkina2021surface}.
Four different silica particles were synthesized to proceed particle assembly in self-lubricating ternary drops, as shown in Figure~\ref{fig:fig4}.
They found the hydrated negatively charged silica particle and statically stabilized silica particles can form supraparticles, as presented in Figures~\ref{fig:fig4} d1 and d2, while dehydrated negatively charged silica particles and positively charged amine-coated particles form flat film-like deposits, as displayed in Figures~\ref{fig:fig4} d3 and d4.
Due to the effects of surface modification on particle, negatively charged PEGylated silica particles, hydrophobicity-enhanced dehydrated silica particles, and positively charged but hydrophobicity-enhanced amine-coated silica particles can adsorb onto the water-oil interface, including the oil ring interface and the oil microdroplets.
As shown in Figures~\ref{fig:fig4}d2-ab, d3-ab, and d4-ab, particle-coated oil microdroplets, the so-called pickering microdroplets, are observed \textit{in suit}  (confocal microscopy: red for particles, and yellow for oil).

It is notable that supraparticles, being micron-scaled assemblies composed of nano-scaled particles, possess a high area-to-volume ratio and are easy to handle at the micro scale.
This combination makes them ideal for catalytic applications, as they offer enhanced catalytic efficiency and practicality in terms of operation and integration.

\begin{figure*}[h]
\centering
\includegraphics[width=0.9\linewidth]{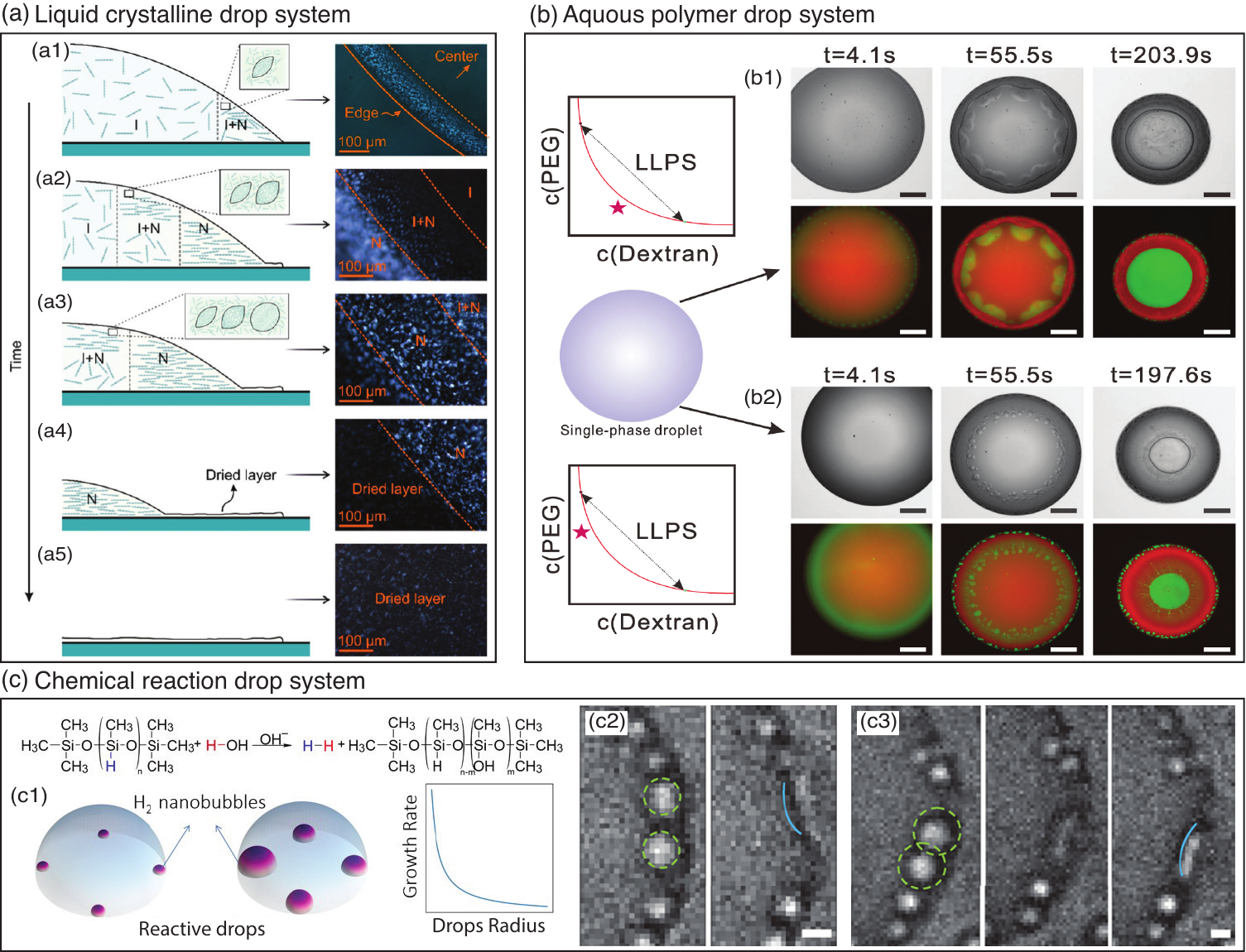}
\caption{Self-lubricating drops of diverse liquid systems.
(a) Evaporating sessile drop of liquid crystalline solution. Evaporation-trigged liquid-liquid crystalline phase separation facilitates the de-pinning behavior of evaporating amyloid fibril-water drop.
Pannel (a) is adapted with permission from ref~\cite{almohammadi2023evaporation}.
(b) Evaporating sessile drop of aqueous two-phase systems. 
The evaporation of Polyethylene Glycol(PEG)/dextran sessile drops induces the non-associative phase separation and the self-lubricating behaviors.
The scale bar is \SI{500}{\micro \meter}.
Pannel (b) is adapted with permission from ref~\cite{guo2021non}.
(c) Reaction between Polymethylhydrosiloxane (PMH) drop and surrounding aqueous media forms H$_2$ nanobubbles at the drop rim. The collapse and coalescence of the bubbles affect the contact line behavior of the drops. 
The scale bar is \SI{1}{\micro \meter}.
Pannel (c) is adapted with permission from ref~\cite{dyett2020accelerated}.
}
\label{fig:fig6}
\end{figure*}

\subsection{Self-lubricating drops for sensing and detection}
\label{sec:s1-2}
Self-lubricating drops enable a novel approach developed for surface-enhanced Raman spectroscopy (SERS) detection, as recently proposed by Dabodiy~\textit{et al.}~\cite{dabodiya2022ultrasensitive}.
The components in the self-lubricating drop include Ag nanoparticles and analytes. 
Facilitated by the self-lubrication effect, analyte-absorbed metal supraparticles can form from the drop evaporation  (Fig.~\ref{fig:fig5}a), and perform as sensitive SERS sites.
These hot spots intensify the SERS signals, leading to an ultra-sensitive detection.
The quantitative analysis of rhodamine 6G is possible with its low concentrations from $10^{-5}$\SI{}{\molar} to $10^{-11}$\SI{}{\molar} (Fig.~\ref{fig:fig5}b).
For a model micro-pollutant in water triclosan, the detection limit is down to $10^{-6}$\SI{}{\molar} (Fig.~\ref{fig:fig5}c).
The ultra sensitive detection mechanism is due to the liquid-liquid extraction during the evaporation of self-lubricating drops.
As demonstrated by Qian~\textit{et al.}~\cite{qian2020one}, nano-extraction enables fluorescence detection limit of nano-molar to pico-molar for fluorescent model compounds in evaporating ternary liquids.
As shown in Figures~\ref{fig:fig6}d and e, although Nile Blue is at a concentration below the detection limit in the initial water-octanol-ethanol mixture, its partitioning into the separated oil (octanol) nano- and micro-droplets gives rise to a fluorescence signal that strong enough to be detected and quantified.

Having observed the impressive enhancements in SERS (Surface-Enhanced Raman Scattering) detection and understanding the underlying principles of extraction, we are convinced that self-lubricating drops have considerable potential for use in sensing and detection across various fields. 
These potential applications include but are not limited to environmental monitoring, food safety, and early diagnosis~\cite{perumal2021towards,huang2020detection,bharati2021flexible, li2022laser}.
Given micro-extraction's broad applicability to diverse multicomponent liquid systems, investigating self-lubricating drops with complex compositions thus becomes particularly  significant for practical use.

\subsection{Self-lubricating drops with complex composition}
The self-lubrication phenomenon occurs not only in the sessile drops of water-alcohol-oil liquid system.
Recent advances have demonstrated the self-lubrication phenomenon happens also with other multicomponent liquids, including liquid crystalline solutions, aqueous polymer solutions, and chemical reaction solutions.

Almohammadi~\textit{et al.} investigated the evaporation-triggered liquid-liquid crystalline phase separation in drops of amyloid fibril-water suspension~\cite{almohammadi2023evaporation}.
The phase separation induces the formation of liquid crystalline tactoids with different shapes and internal structures, as sketched in the inserts of Figure~\ref{fig:fig6}a.
However, they found that the presence of the amyloid fibrils and tactoids resulted in improving the de-pinning behavior of evaporating drop, rather than enhancing the coffee-stain effect. 
Complex dynamics occur in this self-lubricating drop of \textit{liquid crystalline solution}.
As illustrated in Figure~\ref{fig:fig6}a, the rapid evaporation further causes liquid crystalline phase transition in the drop from the isotropic (I) phase,  isotropic+nematic (I+N) phase, to nematic (N) phase.

Aqueous two-phase systems (ATPS) are organic-free and environment-friendly extraction liquid system.
Guo~\textit{et al.} reported the non-associative phase separation triggered by the evaporation of Polyethylene Glycol (PEG)/dextran ATPS sessile drops~\cite{guo2021non}.
The sessile drops with different initial ATPS compositions experience distinctive evaporation processes, as shown in Figure~\ref{fig:fig6}b. 
Their initial compositional points (star symbols) correspond to different evaporative path lines, affecting the phase separation and the consequent drop behaviors.
Nevertheless, the PEG/dextran ATPS sessile drops also exhibit the self-lubricating phenomenon, having smoothly reducing contact area. 

In addition to the evaporation process, recent research by Dyett~\textit{et al.} has explored the potential of using chemical reactions to alter the composition and change contact line behaviors~\cite{dyett2020accelerated}. 
The specific reaction investigated involves the interaction between Polymethylhydrosiloxane (PMH) and water, resulting in the generation of hydrogen.
The experimental results, illustrated in Figure~\ref{fig:fig6}c, revealed that when a PMH sessile drop is immersed in an aqueous medium, nanobubbles of H$_2$  form along the rim of the drop as a byproduct of the reaction.
The collapse and merging of these nanobubbles subsequently impact the behavior of the drop contact line. 
This approach indicates possibilities for the formation of a self-lubricating ring through chemical reactions.

\section{Current opinions and open questions}

The self-lubrication effect occurs in various multicomponent liquid systems with potential applications in extraction and concentration, sensitive sensing, and fabrication of functional materials. 
However, precise control of this effect remains a challenge. 
The multi-physic dynamics involved, as well as their coupling interactions, are complex and fascinating. 
\\

To conclude this review, we summarize the most prominent outstanding questions that remain to be explored.\\

\textit{(1) Interaction between particles and microdroplets:}
Spontaneous nucleated microdroplets can act as templates for advanced regulation.
Manipulating the interaction between laden colloidal particles and the nucleated component microdroplets provides paths for the bottom-to-up fabrication of the multi-scale and functional structures inside the supraparticles, such as tuning the porosity of supraparticles~\cite{liu2019tuning,liu2022controlling}.
These interactions includes capillary interaction and electrostatic interaction in diverse contexts~\cite{banerjee2020drop,spasowka2021chemistry, cui2022functional,shah2022temperature,shim2022diffusiophoresis}.
However, the continuously changing of composition of the self-lubricating drop affects the interaction process, which challenges the manipulation.
\\

\textit{(2) Hydrodynamic effect on particle assembly:}
The self-lubricating drops involve different kinds of flow, including capillary flow~\cite{deegan1997capillary,deegan2000contact}, solutal Marangoni flow~\cite{tan2016evaporation, li2022physicochemical, li2023oil}, and thermal Mrangoni flow~\cite{diddens2017evaporating}. 
These flows interact and cause a chaotic flow field~\cite{diddens2017evaporating, zeng2021periodic,li2023oil, zengh2023}, and the flow field is further coupled with the evolving compositional fields~\cite{lohse2020physicochemical,manikantan2020surfactant}.
Thus, understanding the impact of this physicochemical hydrodynamic process on the particle assembly is essential to the fabrication of the functional micro-/nanostructures~\cite{feng2022long, bindgen2022behavior}.\\ 

\textit{(3) A set of hidden variables from chemical reaction: }
Multicomponent fluids are inevitably accompanied with chemical reactions in many situations~\cite{mathijssen2023culinary,wei2022situ,neto2022we}.
The reaction may create some new component, either miscible or immiscible with the original mixture, which can alter the self-lubricating drop system~\cite{dyett2020accelerated}. 
The chemical reaction variables~\cite{neto2022we} are quite diverse and, thereby, provide self-lubricating drops a broad design space for specified fabrications.
However, since the design options are so huge, constructing a design frame is necessary. \\

\textit{(4) Broad applications from complex composition:}
Aqueous two-phase systems, known as the clean alternatives for traditional organic-water solvent extraction, make self-lubricating drops suitable for bio-engineering processes or environment managements~\cite{wintzheimer2021supraparticles, wu2022nanoextraction, cao2023partitioning}.
It is also notable that liquid crystalline solutions can broaden the applied range for fabrication of semiconductors,  photo-alignment~\cite{dabodiya2022ultrasensitive,bisoyi2016light,abbasi2022chiral,mathijssen2023culinary},

\section*{Declaration of Competing Interest}
The authors declare that they have no known competing financial interests or personal relationships that could have appeared to influence the work reported in this paper.
 \section*{Data availability}
 No data was used for the research described in the article.
 \section*{Acknowledgments}
 H.T. acknowledges the supports from the Natural Science Foundation of China (No. 12102171) and the Natural Science Foundation of Shenzhen (No. 20220814180959001).
D.L. acknowledges supports from the EU (ERC-Advanced Grant Project DDD No. 740479, and ERC-Proof-of-Concept Grant Project No. 862032).
X.Z. acknowledges supports from the Natural Science and Engineering Council of Canada (NSERC) and from the Canada Research Chairs program.


\end{document}